\documentstyle[pra,aps,multicol]{revtex}

\input{epsf}

\tighten  

\begin{document}
\draft
\title{Non-Markovian dynamics in pulsed and continuous wave atom lasers}

\renewcommand{\thefootnote}{\fnsymbol{footnote}}

\author{H. P. Breuer\cite{HPB} , D. Faller\cite{DF} , B. Kappler\cite{BK} and F. Petruccione\cite{FP}}
\address{Albert-Ludwigs-Universit\"at, Fakult\"at f\"ur Physik, \\
         Hermann-Herder Stra{\ss}e 3, D--79104 Freiburg im Breisgau,
         Federal Republic of Germany}
\date{\today}
\maketitle


 \begin{abstract}
The dynamics of atom lasers with a continuous output coupler based
on two-photon Raman transitions is investigated. With the help of
the time-convolutionless projection operator technique the quantum
master equations for pulsed and continuous wave (cw) atom lasers
are derived. In the case of the pulsed atom laser the power of the
time-convolutionless projection operator technique is demonstrated
through comparison with the exact solution. It is shown that in
an intermediate coupling regime where the Born-Markov
approximation fails the results of this algorithm agree with the
exact solution. To study the dynamics of a continuous wave atom
laser a pump mechanism is included in the model. Whereas the pump
mechanism is treated within the Born-Markov approximation, the
output coupling leads to non-Markovian effects. The solution of
the master equation resulting from the time-convolutionless
projection operator technique exhibits strong oscillations in the
occupation number of the Bose-Einstein condensate. These oscillations
are traced back to a quantum interference which is due to the non-Markovian 
dynamics and which decays slowly in time as a result of the dispersion 
relation for massive particles.

\end{abstract}

\begin{multicols}{2}
\narrowtext

\section{Introduction}

Nowadays it is a standard technique to produce a Bose-Einstein
condensate in the laboratory \cite{Ketterle1,Wiemann}.
In order to build a coherent source of atoms, an atom laser, a major
achievement was the coherent extraction of atoms from an atomic trap.
At first a pulsed atom laser was built
\cite{Mewes2,Andrews}, recently also continuous wave atom lasers
have been realized \cite{Hagley,Esslinger}.  A short survey over
the experimental situation is given in \cite{Durfee,Lubkin}.

The theoretical treatment of an atom laser is usually based on the
Born-Markov approximation
\cite{Holland,Zobay,Moore1,Guzman,Wiseman}, which has been used in
quantum optics with great success. However, as it has been shown
recently by Moy and coworkers \cite{Moy1}, this approximation
fails in a realistic parameter regime. In this article we will
outline a different approach, which is based on the
time-convolutionless (TCL) projection operator technique
\cite{Chaturvedi,Fumiaki1,Kappler1} to study non-Markovian effects resulting
from the output coupling. This technique is based on a
perturbative expansion in powers of the output coupling strength.

The master equation resulting from the perturbative expansion has a
form similar to the Born-Markov master equation and is also local
in time. This makes the equation of motion easy to solve.
As we will see the second order
perturbation theory corresponds to the Born-Markov approximation.
By taking higher orders of the expansion non-Markovian effects can
be studied in a systematic way.
It is demonstrated that the perturbative expansion  holds in an
intermediate coupling regime, which corresponds to realistic
parameters, while the Born-Markov approximation fails for these
parameters.

This paper consists of two parts and is structured as
follows. In section~\ref{pulsed_atom_lasers} we investigate the
validity of the time-convolutionless projection operator technique
(TCL) for a pulsed atom laser.
The model investigated was  discussed by Moy {\it et al.}
\cite{Moy1}. They showed that the Born-Markov approximation fails
for realistic parameters. We demonstrate that the
time-convolutionless projection operator technique agrees with the
exact solution for these parameters. Afterwards we apply the TCL
algorithm to a simple model of a continuous wave atom laser in
section~\ref{cw_al}. The numerical results obtained from a
simulation using a perturbation expansion including 4th order
clearly reveal strong oscillations in the occupation number of the
Bose-Einstein condensate (BEC). As will be shown these oscillations can
be interpreted as a quantum interference effect which clearly reveals
departures from the golden rule and demonstrating the non-Markovian dynamics of the
atom laser. A short summary concludes the article.

\section{Pulsed atom lasers} \label{pulsed_atom_lasers}

In this part we will investigate the dynamics of a pulsed atom
laser. This means that there is an output
coupling mechanism which transfers the atoms out of the trap but
there is no pump mechanism which supplies the BEC with new atoms.
When the trap is empty it is replenished and a new cycle starts.
Therefore our starting point is an existing BEC inside an atomic
trap and we study only the output coupling.

This part is organized as follows. We briefly introduce our atom laser model
and derive the exact equation of motion for the expectation value
of the atomic number operator of the BEC in section~\ref{ex_sol}.
In section~\ref{b_m_sec} we discuss the Born-Markov approximation.
The TCL-algorithm is then applied to the model in
section~\ref{tcl_sec}. Afterwards the numerical results are
discussed in section~\ref{num_res}. A model of a continuous wave
atom laser which includes a pumping mechanism to compensate
losses by the output coupling will be introduced in part~\ref{cw_al}.

\subsection{Exact solution} \label{ex_sol}

Usually, the output coupling from
a BEC to a large reservoir is described within the Born-Markov
approximation. This description is based upon a master equation of the form
\begin{equation}
 \frac{d}{d t} \rho_{\text{sys}}(t) =  \gamma_{M} \;
    { \cal D} \lbrack a \rbrack \, \rho_{\text{sys}}(t) \; ,
\end{equation}
where the single trap mode occupied by the BEC is described by the
creation and annihilation operators $a^{\dagger}, a$ and
$\gamma_{M}$ is the Markovian decay rate. The superoperator ${\cal
D}$ is of Lindblad form \cite{Lindblad} and is defined by
\begin{equation}
 {\cal D } \lbrack a  \rbrack\,  P = a P a^{\dagger} - \frac{1}{2} (a^{\dagger} a P + P a^{\dagger} a) \;.
\end{equation}
In the Markovian approach one obtains a master equation containing
only system variables. Two basic assumptions underlying the
Born-Markov approximation are that (i) if the system and the reservoir
are uncorrelated at the beginning they remain uncorrelated for
later times and (ii) that the evolution of the system is Markovian. The
Markovian property means that the future evolution of the system
depends only on the present state and  is independent of
the previous history of the system. As it is shown in Ref. \cite{Moy1} the
Born-Markov approximation is not valid in a realistic parameter
regime of atom lasers.

We want to investigate a Raman output coupler through state change, which was  suggested by Moy
{\it et al.} \cite{Moy1,Moy2,Moy3,Hope2}. Two lasers are
tuned to a two-photon resonance to couple an initial atomic state
inside the trap to a final atomic state outside the trap. Due to
the conservation of linear momentum the atoms receive momentum
kicks during the Raman transitions which pushes the atoms out of
the trap.
If we assume that the lasers are detuned far from single photon
resonances, then all initially empty modes of the atomic trap
remain empty and we can neglect all modes  besides the ground
state mode occupied from the BEC.\@ In addition, we also ignore the
effects of atom-atom interactions. The resulting Hamiltonian $H$ is of the form
\begin{mathletters}
\label{ham_ops}
\begin{eqnarray}
  H &=&  H_{\text{sys}} + H_{\text{res}}  \label{hges}
               + H_{\text{int}} ,\\
         H_{\text{sys}} &=& \hbar \omega_{0} \, a^{\dagger} a ,  \\
         H_{\text{res}} &=& \int_{-\infty}^{+\infty} dk \, \hbar
         \omega_{k} \; b_{k}^{\dagger} b_{k} , \\
         H_{\text{int}} &=& -i \hbar \, ( B a^{\dagger} - B^{\dagger} a ) \;,\label{eins}
\end{eqnarray}
\end{mathletters}
where we have defined
\begin{equation}
 B =  \int_{-\infty}^{+\infty} dk \, \kappa(k)
 b_{k}\; , \quad  \quad \omega_{k} = \frac{\hbar k^{2}}{2 M}\;.   \label{B_formel}
\end{equation}
The bosonic creation and annihilation operators for the free
atomic state are denoted by $b_{k}^{\dagger},b_{k}$. The ground state trap
energy is denoted by $\hbar \omega_{0}$ and $\hbar \omega_{k}$ is the
energy of the free output atomic state, $M$ is the atomic mass.
The function $\kappa(k)$ describes the strength and spectral form of the coupling. In
the specific case of a Raman output coupler and a harmonic trap
with a Gaussian ground state~\cite{Moy3,Hope2} of width $\sigma_{k}$ in $k$-space
given by
\begin{equation}
 \Psi(k) = (2 \pi \sigma_{k})^{-1/4}\; \exp(-k^{2}/(4
 \sigma_{k}^{2}))\; ,
\end{equation}
one obtains together with Eq. (\ref{eins}) for the interaction
Hamiltonian $H_{\text{int}}$
\begin{equation}
 \kappa(k) = \frac{i\, \Gamma^{1/2}}{\sqrt[4]{2 \pi \sigma_{k}^{2}}}
    \exp(- (k-k_{0})^{2}/( 4 \sigma_{k}^{2}) ) \;.  \label{kappa_ausdr}
\end{equation}
Here $\Gamma^{1/2}$ denotes the coupling strength of the output
coupling and $\hbar k_{0}$ is the momentum transferred through the
Raman transition. 
For the sake
of simplicity we assume that we have two counter propagating laser
beams with the wave vectors $k_{L1} \approx - k_{L2}$. Therefore we
have $k_{0} \approx 0$. With the atomic dispersion relation from
Eq. (\ref{B_formel}) one derives the spectral density $J(\omega)$ of the output
coupling strength,
\begin{equation}
J(\omega) = \frac{\Gamma}{\sqrt{\pi \alpha \omega}} \;e^{-\omega/\alpha}\;,\label{sp_den}
\end{equation}
which diverges at $\omega=0$. The quantity $\alpha$ is given by
\begin{equation}
\alpha =  \frac{\hbar \sigma_{k}^{2}}{2 M} \;.
\end{equation}
With the help of the function $J(\omega)$ we can
deduce the Heisenberg equations of motion for the system operators
$a^{\dagger}$ and $a^{\dagger} a$. Assuming that the reservoir is empty at
the beginning, $\langle b_{k}^{\dagger} b_{k} \rangle(0)=0$, we obtain
\begin{mathletters}
\label{heis_eq}
\begin{eqnarray}
   \frac{d}{dt} \langle a^{\dagger}(t)\rangle &=& i \omega_{0} \langle
a^{\dagger}(t)\rangle - \nonumber  \\  & &{}
   \int_{0}^{t} d\tau
   \; f^{*}(\tau) \langle a^{\dagger}(t-\tau) \rangle e^{i \omega_{0} \tau} , \label{gl_a}
   \\
   \frac{d}{dt} \langle a^{\dagger}(t) a(t) \rangle &=& - \int_{0}^{t} d\tau
   \; f^{*}(\tau) \langle a^{\dagger}(t) \; a(t-\tau)\rangle e^{- i \omega_{0} \tau}
   \nonumber \\  & &{}   + \textrm{h.c.} \label{gl_aa}
\end{eqnarray}
\end{mathletters}
The function
\begin{eqnarray}
 f(\tau) &=& \text{Tr} \lbrace \rho_{\text{res}}B(\tau) B^{\dagger}(0) \rbrace e^{i \omega_{0} \tau} \\
         &=& \int_{0}^{\infty} d\omega \; J(\omega) e^{i  (\omega_{0}-\omega) \tau}\\
          &=&{} \frac{e^{i \omega_{0} \tau} \Gamma}{\sqrt{1+i \alpha \tau}}\; , \label{f_tau}
\end{eqnarray}
is the reservoir correlation function apart from a factor
$e^{i \omega_{0} \tau} $.

The formal solution of the integro-differential equations
(\ref{heis_eq}) can  be expressed in terms of inverse
Laplace transforms. However, the numerical evaluation of the inverse
Laplace transforms is very difficult \cite{Hope1}. As is easily shown, the solution of Eq.
(\ref{gl_aa}) can be written as
\begin{equation}
  \langle a^{\dagger}\, a \rangle(t) =  c(t)\;c^{*}(t)\;.
\end{equation}
where $c^{*}(t)$ is a solution of Eq. (\ref{gl_a}) with the initial
value $c(0) = \sqrt{\langle  a^{\dagger}\, a \rangle(0)}$.
Hence one solves Eq. (\ref{gl_a}) through direct numerical
integration and obtains the expectation value of the atom number
by taking the squared absolute value.

\subsection{Born-Markov approximation} \label{b_m_sec}

Now we apply the Born-Markov approximation to the above model of
an atom laser. The Born-Markov approximation consists in making
the Born approximation that is, one assumes that system and reservoir are uncorrelated
initially and that they remain so for later times. Hence
the total density operator $ \rho_{\text{tot}}(t)$ can be written as
\begin{equation}
 \rho_{\text{tot}}(t) = \rho_{\text{sys}}(t) \otimes \rho_{\text{res}}(0)\;,
\end{equation}
where $\rho_{\text{sys}}(t)$ denotes the system density operator.
In addition, it is assumed here that the reservoir density operator
does not change in time $ \rho_{\text{res}}(t) = \rho_{\text{res}}(0)$.
The Markov approximation is based on the
assumption that the time scale $t_{\text{sys}}$, which describes
the relaxation of the reduced system, is much greater than the
time scale $t_{\text{res}}$ which represents a measure for the width
of the reservoir correlation function, i. e.,
\begin{equation}\label{timescale}
t_{\text{sys}} \gg t_{\text{res}}\;.
\end{equation}
The time scale $t_{\text{res}} $ is obtained from the reservoir
correlation function $f(\tau)$. Because of the slow decaying
shape of $f(\tau)$ in Eq. (\ref{f_tau}) the atom laser shows a
strong non-Markovian behavior. A more detailed discussion of the
application of the Born-Markov approximation to atom lasers can be
found in Ref. \cite{Moy1,Jack}. Within the Born-Markov approximation we get
from Eqs. (\ref{ham_ops})  the following master equation:
\begin{eqnarray} \label{m_eq}
  \frac{\partial}{\partial t} \rho_{\text{sys}}(t) &=& - \frac{i}{2} S_{M}  [a^{\dagger} a,
  \rho_{\text{sys}}(t)] + \gamma_{M}  \left\{
   - \frac{1}{2} a^{\dagger} a \rho_{\text{sys}}(t) \nonumber
 \right. \\ & &{} \left. - \frac{1}{2} \rho_{\text{sys}}(t)
  a^{\dagger} a +    a \rho_{\text{sys}}(t) a^{\dagger} \right\}\;.
\end{eqnarray}
Here, the Markovian Lamb shift $S_{M}$ and the Markovian decay
rate $\gamma_{M}$ take the form
\begin{equation} \label{gam_mar}
 \gamma_{M}=\int_{0}^{+\infty} dt \;\phi(t) \; , \quad
S_{M}=\int_{0}^{+ \infty} dt \; \psi(t) \;.
\end{equation}
The functions $\phi(t)$ and $\psi(t)$ are proportional to the real and
imaginary part of $f(t)$,
\begin{equation}   \label{phi_psi}
 f(t) = \frac{1}{2} (\phi(t) + i\, \psi(t))\;.
\end{equation}
From the master equation (\ref{m_eq}) we easily get the
expectation value of the atom number operator $a^{\dagger} a$
\begin{equation}
   \langle a^{\dagger} a\rangle(t) = \langle a^{\dagger} a\rangle(0) \; e^{- \gamma_{M}t}\;. \label{BM}
\end{equation}

The above results enable one to derive from inequality
(\ref{timescale}) an explicit condition for the Born-Markov
approximation. The system time scale $t_{\text{sys}}$ is given by
\begin{equation}
 t_{\text{sys}} = \frac{1}{\gamma_{M}} = \sqrt{\frac{\omega_{0}
\alpha}{4 \pi}} \frac{e^{\omega_{0}/ \alpha}}{\Gamma} \;.
\end{equation}
This result is easily obtained by performing the first integral in
(\ref{gam_mar}). As discussed in \cite{Moy1} we define
$t_{\text{res}}$ as the half-width of the integral of the 
real part of $f(\tau)$. The resulting equation is solved
numerically. In the considered parameter regime one obtains $t_{\text{res}}
\approx 0.4/\omega_{0}$. Hence the time scale condition
\begin{equation} \label{timescale2}
 \frac{t_{\text{sys}}}{t_{\text{res}}} \approx  \omega_{0}^{3/2} \sqrt{\frac{\alpha}{\pi}}
                                          \frac{e^{\omega_{0}/\alpha}}{\Gamma}  \gg 1
\end{equation}
must be satisfied for the Born-Markov approximation to be valid.

In our simulations we take similar parameters as discussed in
\cite{Moy1}. The atom mass is $ M \approx 2 \times 10^{-26}\;
\mathrm{kg}$. Realistic parameters for the system frequency
$\omega_{0}$ and the coupling strength $\Gamma$ are $\omega_{0}/(2
\pi) \approx 123\;\mathrm{s}^{-1}$, $\Gamma \approx 10^{5} \;
\mathrm{s}^{-2}$ \cite{Andrews,Hagley,Esslinger,Moy3,Mewes1}. The
standard deviation $\sigma_{k}$ in $k$-space of the coupling
function $\kappa(k)$ is assumed to be $\sigma_{k} \approx 10^{6}
\mathrm{m}^{-1}$ corresponding to a wave length $\lambda \approx 2
\mu\! \mathrm{m}$.

The only parameter which is varied is the coupling strength
$\Gamma$, thereby we can systematically change the ratio of the
time scale condition (\ref{timescale2}). In Fig.~\ref{fig2} we
depict the normalized expectation value of the atomic number
operator obtained from the exact solution and from the Born-Markov
approximation for $\Gamma = 5 \times 10^{4} \mathrm{s}^{-2}$
leading to a ratio of time scales $t_{\text{sys}}/t_{\text{res}} =
16 $.
Obviously the Born-Markov approximation is not valid in this
parameter regime. Attenuating the coupling constant $\Gamma$ about
a factor of $5$ suffices to reach parameters where the Born-Markov
approximation holds.

In the following section we apply the time-convolutionless
projection operator technique to our model. This leads to a
perturbative expansion for the master equation, which allows us to
analyze the behavior of an atom laser in the non-Markovian
parameter regime. As we will see in section~\ref{num_res} the
solution of this master equation agrees very well with the exact
solution for the parameters mentioned above.

\subsection{Time-convolutionless projection operator technique}\label{tcl_sec}

It is usually assumed that a non-Markovian dynamics necessarily
involves a density matrix equation containing a time-convolution
kernel. However, employing the time-convolutionless projection
operator technique we can derive  an equation of motion for the
system density operator which is local in time. The following
discussion briefly summarizes this approach. A more detailed
discussion can be found in
\cite{Chaturvedi,Fumiaki1,Kappler1}. Especially  a discussion of the range of validity
of this approach is given  in Ref. \cite{Kappler1}. 

The starting point is the Liouville-von Neumann equation for the
density matrix of the total system $\rho_{\text{tot}}$
\begin{equation} \label{neum}
  \frac{\partial}{\partial t} \rho_{\text{tot}}(t) = -i \alpha \lbrack
  H_{\text{int}}(t) , \rho_{\text{tot}}(t) \rbrack \equiv
\alpha L(t) \rho_{\text{tot}}(t)\;.
\end{equation}
Here $\alpha$ denotes the strength of the coupling, and
$ H_{\text{int}}(t)$ is the interaction Hamiltonian in the
interaction picture. If we define the projection operator ${\cal P}$
\begin{equation}
   {\cal P} \rho_{\text{tot}}(t) = \text{Tr}_{\text{res}} \lbrace
\rho_{\text{tot}}(t) \rbrace \otimes \rho_{\text{res}}
   = \rho_{\text{sys}}(t) \otimes \rho_{\text{res}} \; ,
\end{equation}
it is possible to deduce an equation of motion for the system density
operator $ \rho_{\text{sys}}(t)$
\begin{equation}
    \frac{\partial}{\partial t} {\cal P}  \rho_{\text{tot}}(t) =
K(t) {\cal P}  \rho_{\text{tot}} (t)  \label{tcl_gl}\;.
\end{equation}
Here $\text{Tr}_{\text{res}}$ denotes the trace over the reservoir.
Under certain conditions which are always satisfied for short times and in the
weak coupling regime (see e.g. Ref. \cite{Kappler1}) it is possible to derive a perturbative expansion for the
generator $K(t)$ of the time-convolutionless master equation in powers of the
coupling strength $\alpha$
\begin{equation}   \label{entw}
   K(t) = \alpha^{2} K^{(2)}(t) +  \alpha^{4} K^{(4)}(t) + \alpha^{6} K^{(6)}(t)
           + \dots \;.
\end{equation}
The superoperators $K^{(n)}(t)$ are given by
\begin{equation} \label{entw2}
  K^{(n)}(t) = \int \limits_{0}^{t} dt_{1} \int \limits_{0}^{t_{1}}
  dt_{2} \ldots \int \limits_{0}^{t_{n-2}} dt_{n-1}  k_{n}(t,t_{1}
  \ldots ,t_{n-1}) \; ,
\end{equation}
where
\begin{eqnarray}    \label{cum}
  k_{n}(t,t_{1}, \ldots ,t_{n-1}) &=& \sum (-1)^{q-1}   {\cal P}
  L(t) \dots L(t_{i}) \nonumber \\ & &{}{\cal P} L(t_{j}) \dots L(t_{l}) {\cal P}
  L(t_{m}) \dots\;.
\end{eqnarray}
The sum in (\ref{cum}) is to be taken over all possible insertions of  ${\cal P}$'s
in between the $n$ factors L, while keeping the chronological order in
time  between two successive insertions of ${\cal P}$
\cite{Fumiaki1}. This means  $t > \dots > t_{i}$ and $t_{j} > \ldots >  t_{l}$.
The constant $q$ is the number of inserted
projection operators ${\cal P}$'s.
Assuming that all odd moments of $H_{\text{int}}(t)$ with respect to $\rho_{\text{res}}$
vanish we have ${\cal P} L(t_{1}) \dots L(t_{2k+1}) {\cal P} = 0$. Therefore, all terms
containing odd powers of the coupling strength $\alpha$ disappear in equation (\ref{entw}).
With the help of Eqs. (\ref{entw2}) and (\ref{cum}) we obtain the superoperators
\begin{equation}
  K^{(2)}(t) = \int_{0}^{t} dt_{1} \; {\cal P} L(t) L(t_{1})
  {\cal P} \; ,
\end{equation}
and
\begin{eqnarray}
  &K&^{(4)}(t) = \int_{0}^{t} dt_{1}  \int_{0}^{t_{1}}  dt_{2} \label{k4}
 \int_{0}^{t_{2}} dt_{3}   \\
 \lefteqn{\times \lbrace {\cal P} L(t) L(t_{1}) L(t_{2}) L(t_{3}) {\cal P}  -
 {\cal P} L(t) L(t_{1}) {\cal P} L(t_{2}) L(t_{3}) {\cal P}} \nonumber \\
 &-& {\cal P} L(t) L(t_{2}) {\cal P} L(t_{1}) L(t_{3}) {\cal P} - {\cal P} L(t)
  L(t_{3}) {\cal P} L(t_{1}) L(t_{2}) {\cal P} \rbrace \;. \nonumber
\end{eqnarray}
We have determined also the sixth order superoperator $K^{(6)}$. Because of its
length the expression for $K^{(6)}(t)$ is not presented
here: it is  a fivefold integral of 45 terms, each containing 6 superoperators $L(t)$. 

The above superoperators can be expressed in terms of commutators whose
evaluation is simplified by invoking the bosonic commutation
relations. For the present study we have determined
$K^{(2)}$, $K^{(4)}$, and $K^{(6)}$ with the help of the computer
algebra system Mathematica. This system enables the evaluation of the
expansion  coefficients $K^{(n)}(t)$.

Transforming Eqs. (\ref{eins}) and (\ref{B_formel}) to the interaction picture,
the interaction Hamiltonian $H_{\text{int}}(t)$ reads
\begin{equation}
   H_{\text{int}}(t) = - i \hbar\; \lbrace B(t) a^{\dagger}(t) - B^{\dagger}(t) a(t) \rbrace \; ,
\end{equation}
where
\begin{equation}
  B(t) = \int_{- \infty}^{\infty} d k \; \kappa(k) b_{k} e^{-i \omega_{k} t}\; , \quad
  a(t) = a\; e^{-i \omega_{0} t} \;.
\end{equation}
The expansion parameter in the case of the atom laser is the
coupling strength $\Gamma^{1/2}$. With the assumption of an
initially empty Gaussian reservoir, $\text{Tr}_{\text{res}}
\lbrace \rho_{\text{res}}B^{\dagger}(t) B(t_{1}) \rbrace=0$, we
obtain a master equation in Lindblad form
\begin{eqnarray} \label{tcl_mgl}
   \lefteqn{\frac{\partial}{\partial t} \rho_{\text{sys}}(t) = - \frac{i}{2} S(t)
   \lbrack a^{\dagger} a , \rho_{\text{sys}}(t) \rbrack  \nonumber} \\
   &+& \frac{1}{2} \gamma(t) \left\{
   - a^{\dagger} a \rho_{\text{sys}}(t) - \rho_{\text{sys}}(t) a^{\dagger}
   a + 2  a \rho_{\text{sys}}(t) a^{\dagger} \right\} \;.
\end{eqnarray}
Here the  Lamb shift $S(t)$ and the  decay rate $\gamma(t)$ are given by
\begin{eqnarray}
 S(t) &=& S^{(2)}(t)+S^{(4)}(t)+S^{(6)}(t) \; , \\
 \gamma(t) &=& \gamma^{(2)}(t)+\gamma^{(4)}(t)+\gamma^{(6)}(t) \;.
\end{eqnarray}
The functions $S^{(n)}(t)$ and $\gamma^{(n)}(t)$ can be easily  evaluated as
integrals over the known functions $\phi(t), \; \psi(t)$ [see Eq. (\ref{phi_psi})].
We find
\begin{mathletters}
\begin{eqnarray}
 S^{(2)}(t) &=& \int_{0}^{t} dt_{1} \psi(t-t_{1}) \; ,\label{s2} \\
 S^{(4)}(t) &=&   \frac{1}{2} \int_{0}^{t} dt_{1}
                  \int_{0}^{t_{1}} dt_{2} \int_{0}^{t_{2}} dt_{3}\;
                   \lbrack \psi(t-t_{2}) \phi(t_{1}-t_{3}) \nonumber \\ & &{}
  +\phi(t-t_{2}) \psi(t_{1}-t_{3})  + \psi(t-t_{3}) \label{s4}
 \phi(t_{1}-t_{2}) \nonumber \\ & &{} +\phi(t-t_{3}) \psi(t_{1}-t_{2})
\rbrack\; ,
\end{eqnarray}
\end{mathletters}
and
\begin{mathletters}
\begin{eqnarray}
  \gamma^{(2)}(t) &=& \int_{0}^{t} dt_{1} \phi(t-t_{1}) \label{ga_tcl2}\; , \\
  \gamma^{(4)}(t) &=&
 \frac{1}{2}  \int_{0}^{t} dt_{1}  \int_{0}^{t_{1}} dt_{2}
\int_{0}^{t_{2}}  dt_{3} \; \lbrack 
  \phi(t-t_{2}) \phi(t_{1}-t_{3}) \nonumber \\
  & &{} + \phi(t-t_{3}) \phi(t_{1}-t_{2}) -  \psi(t-t_{3})  \psi(t_{1}-t_{2}) \nonumber \\  
  & &{} - \psi(t-t_{2}) \psi(t_{1}-t_{3})  \rbrack\;. \label{g4}
 \end{eqnarray}
\end{mathletters}
Again we refrain from presenting the sixth order terms because of
their length, but they are also given as simple integrals.

Note that we have derived a  master equation in Lindblad form
without making the Born-Markov approximation. If we keep only the
second order terms and extend the range of integration in Eqs.\
(\ref{s2}) and (\ref{ga_tcl2}) to infinity we get the Born-Markov
master equation (\ref{m_eq}). Hence our master equation has the
same simple form as equation (\ref{m_eq}), but the parameter
regime in which Eq. (\ref{tcl_mgl}) is valid is much larger. This
will be demonstrated in section~\ref{num_res}.

From Eq. (\ref{tcl_mgl}) one derives the equation of motion for
the occupation number of the BEC.\@ We are only interested in the
normalized occupation number $n(t)=\langle a^{\dagger}
a\rangle(t)/\langle a^{\dagger} a \rangle(0)$. One easily
demonstrates that  $n(t)$ is given by
\begin{equation}
  n(t) = \exp\left(- \int_{0}^{t} \gamma(t_{1}) dt_{1}\right)\;.
\end{equation}
Hence it suffices to evaluate the integrals over the time
dependent decay rates $\gamma^{(n)}(t)$ in the different orders of
the coupling strength $\Gamma^{1/2}$ to get the normalized occupation
number of the BEC.\@ Provided $\gamma^{(n)}(t)$ is positive the quantity $F(t) = 1 - n(t)$ can be interpreted
as the waiting time distribution  \cite{Carmichael,BP1} for a transition of
a single atom out of the trap.

\subsection{Numerical results} \label{num_res}

In our simulation we take the parameters as mentioned at the end of section~\ref{b_m_sec}.
Through variation of the coupling strength $\Gamma$ we
will go beyond  the parameter regime in which the Born-Markov approximation
is valid and compare the results obtained from the time-convolutionless projection operator technique with
the exact solution.

\begin{figure}
\begin{minipage}[b]{0.46\linewidth}
\centering\epsfxsize\linewidth\epsffile{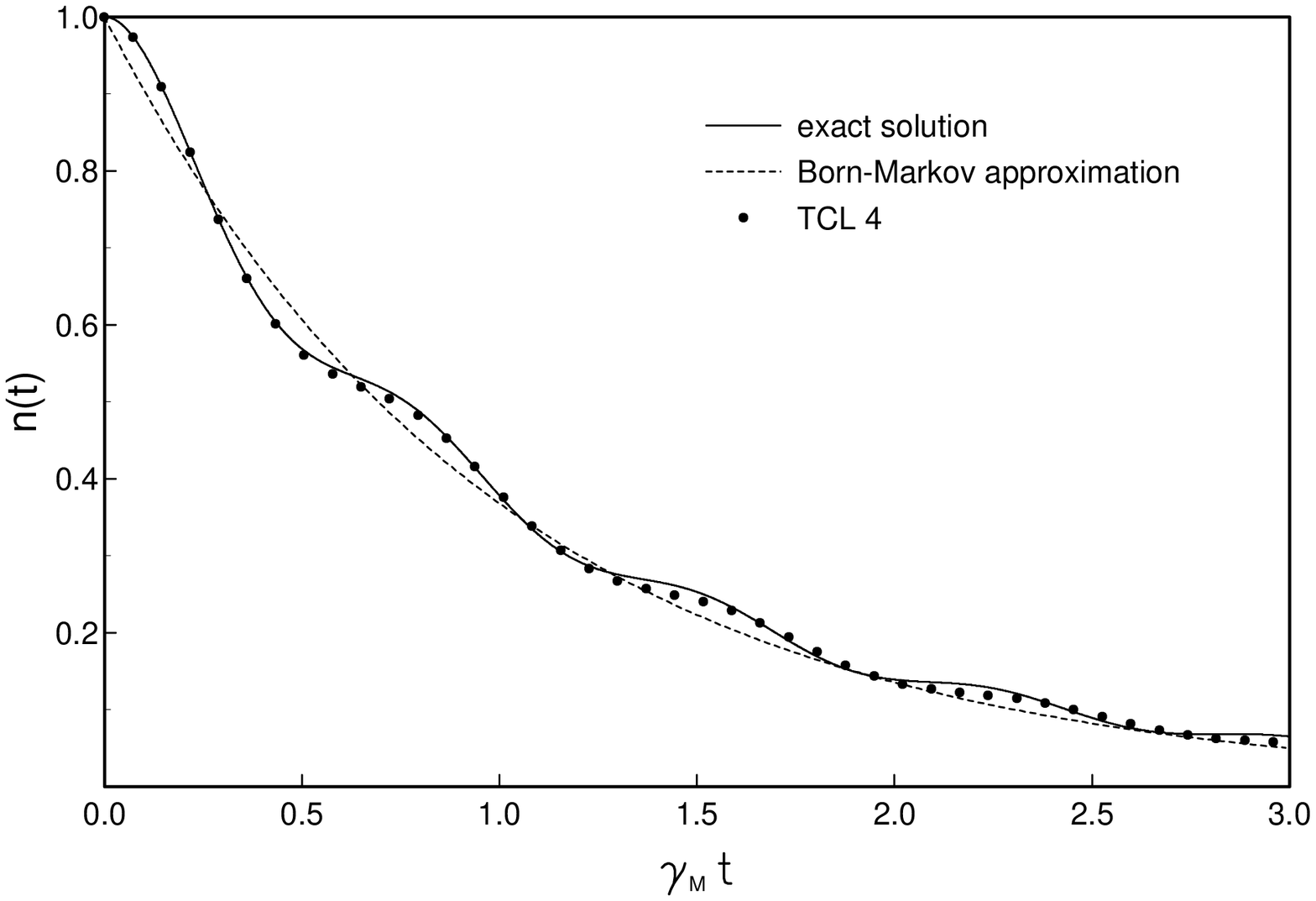}
  \caption{Normalized occupation number $n(t) = \langle a^{\dagger} a\rangle(t) / \langle a^{\dagger} a \rangle(0)$
           of the BEC in the atomic trap. Plotted
           are Born-Markov approximation, exact solution and perturbation
           expansion to 4th order in the coupling strength $\Gamma = 5 \times 10^{4} s^{-1}$.
           This leads to a ratio of time scales $t_{\text{sys}}/t_{\text{res}} = 16$. }
  \label{fig2}
\end{minipage}\hfill
%
\begin{minipage}[b]{0.46\linewidth}
\centering\epsfxsize\linewidth\epsffile{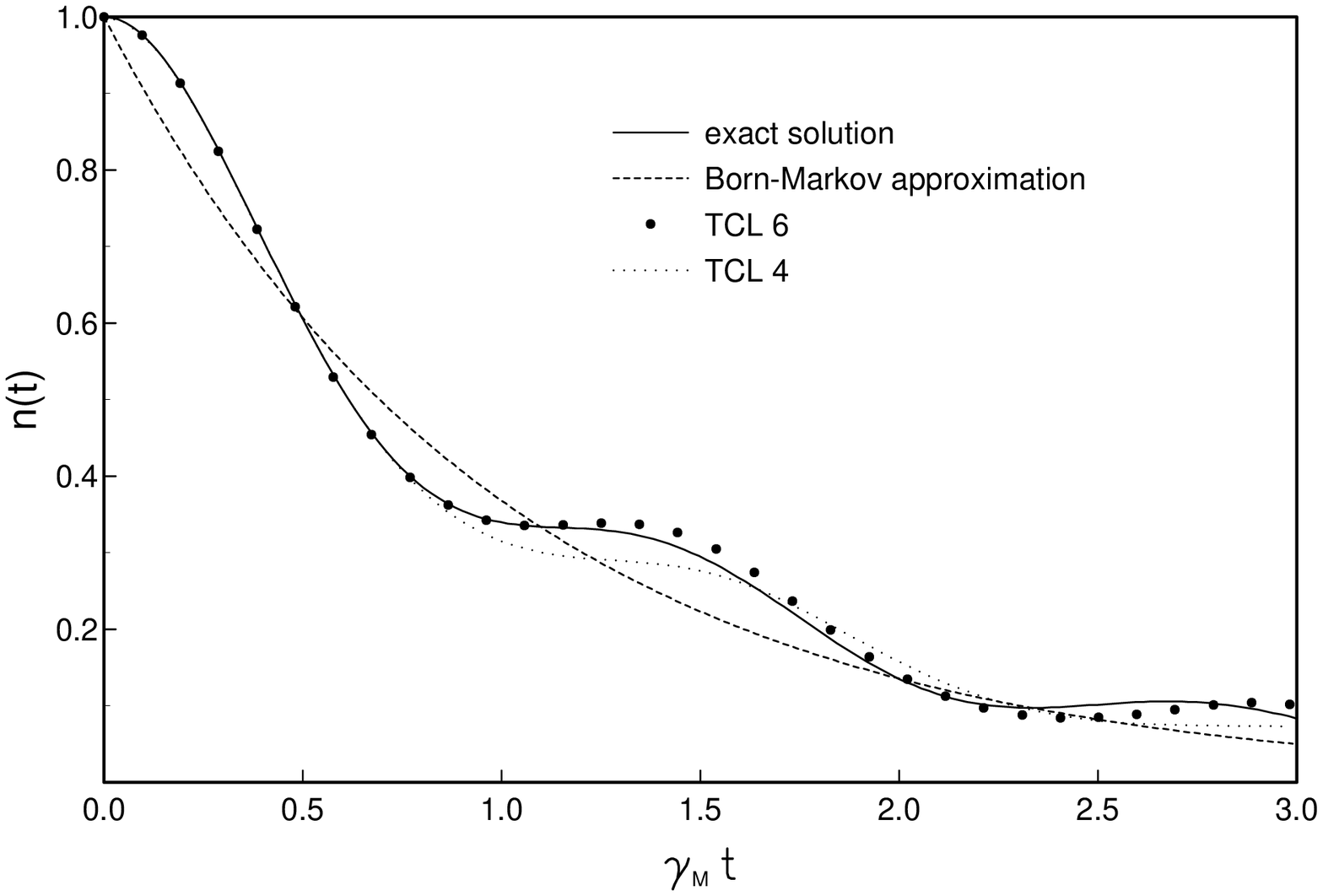}
  \caption{The normalized occupation number
           $n(t) = \langle a^{\dagger} a\rangle(t) / \langle a^{\dagger} a \rangle(0)$
           of the BEC: exact results and the
           results from the Born-Markov approximation, resp., from 4th and 6th order
           perturbation theory. The parameter $\Gamma = 10^{5} s^{-1}$ is chosen such that the ratio of time scales becomes
           $t_{\text{res}}/t_{\text{sys}} = 8$. }
  \label{fig3}
\end{minipage}
%
\begin{minipage}[b]{0.46\linewidth}
 \centering\epsfxsize\linewidth\epsffile{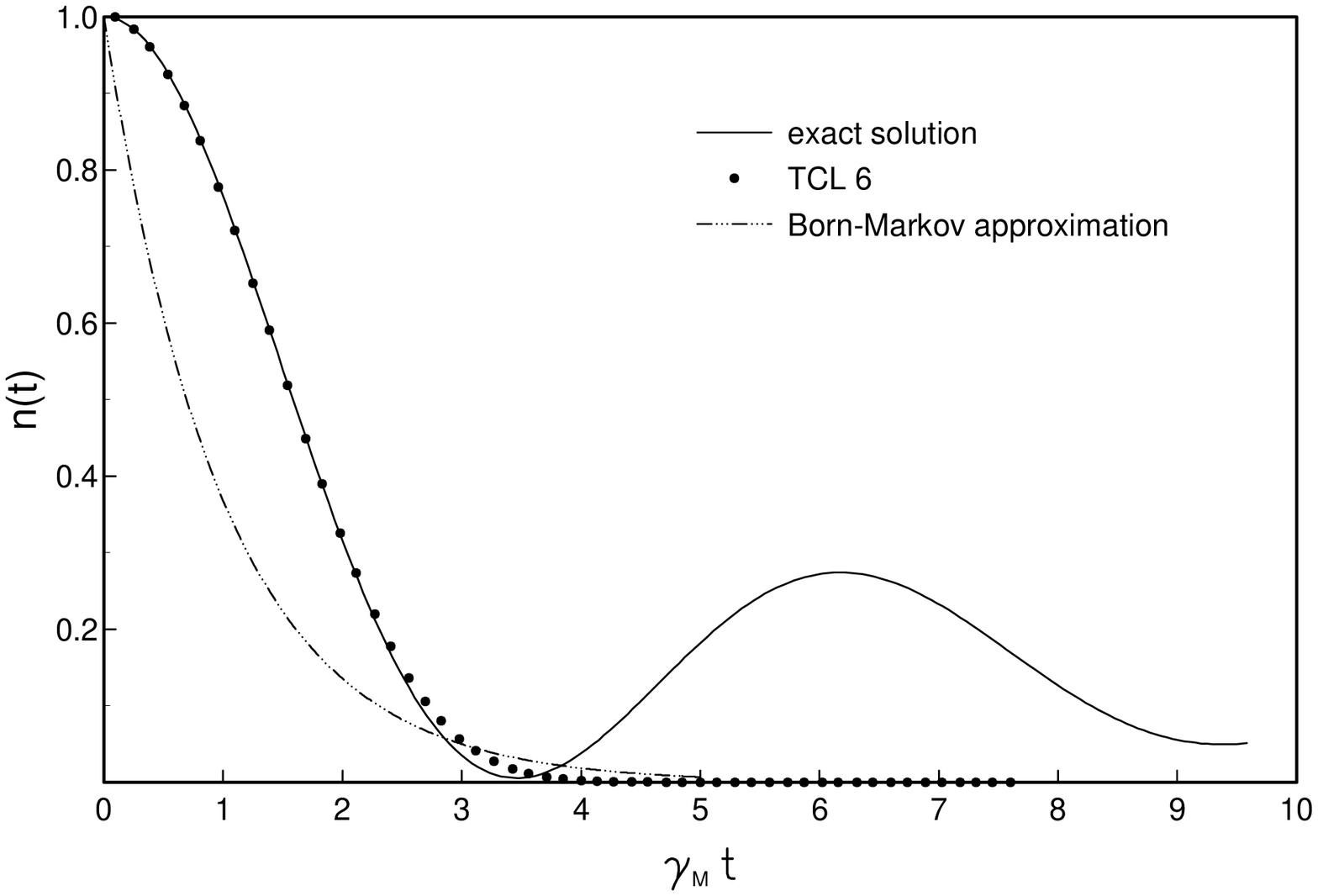}
  \caption{Normalized occupation number
           $n(t) = \langle a^{\dagger} a\rangle(t) / \langle a^{\dagger} a \rangle(0)$
           of the BEC: Born-Markov approximation, exact solution and perturbation
           expansion to 6th order in the coupling strength  $\Gamma = 10^{6} s^{-1}$.
           This leads to a ratio of time scales  $t_{\text{res}}/t_{\text{sys}} = 0.8$. }
  \label{fig4}
\end{minipage}\hfill
%
\begin{minipage}[b]{0.46\linewidth}
  \centering\epsfxsize\linewidth\epsffile{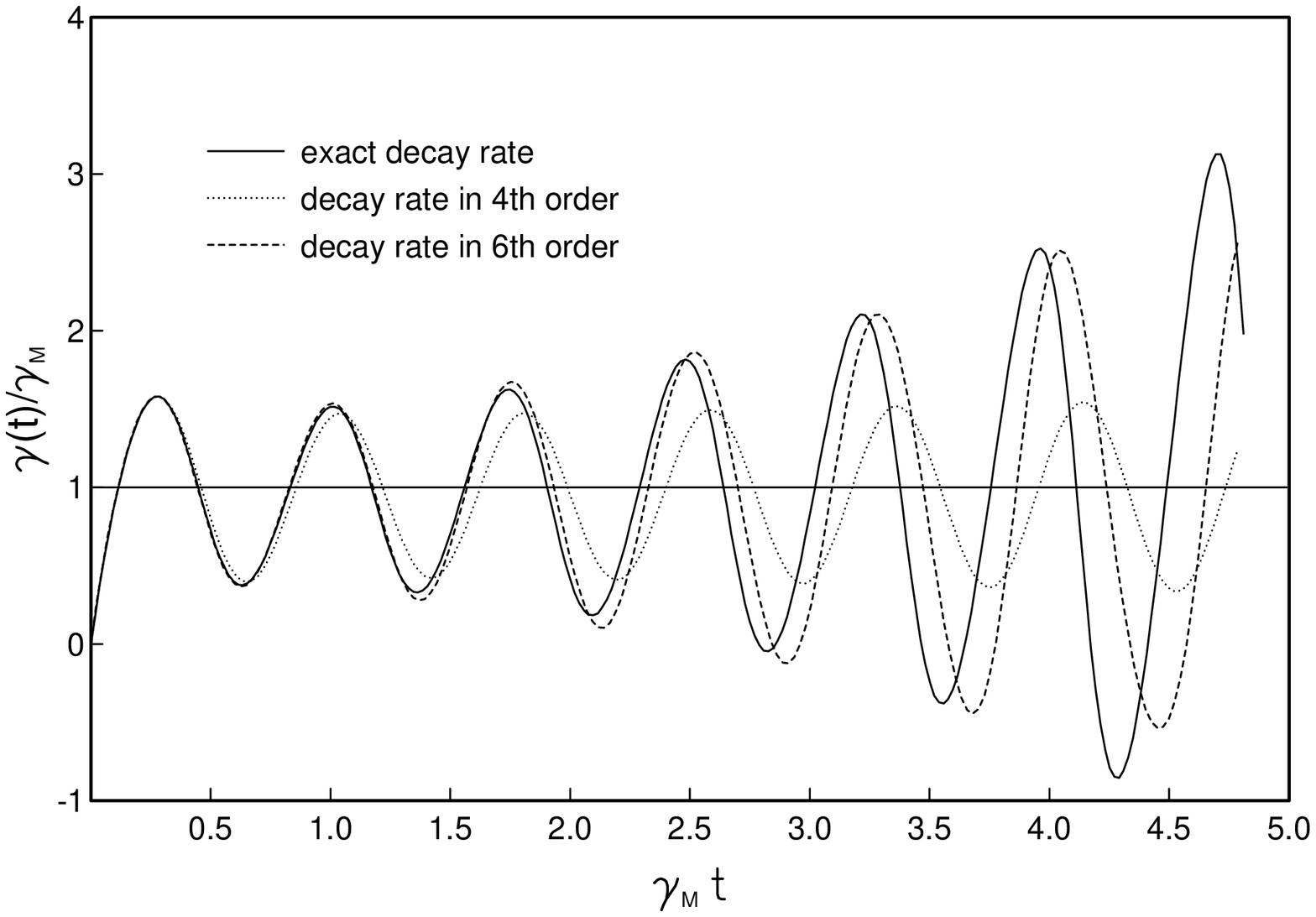}
  \caption{Comparison of decay rates. Plotted are Born-Markov, exact
           decay rates and the decay rates in 4th and 6th order perturbation
           theory. The value of the coupling strength is  $\Gamma = 5\times 10^{4} s^{-1}$.     }
  \label{fig5}
\end{minipage}
%
\noindent
\begin{minipage}[b]{0.46\linewidth}
  \centering\epsfxsize\linewidth\epsffile{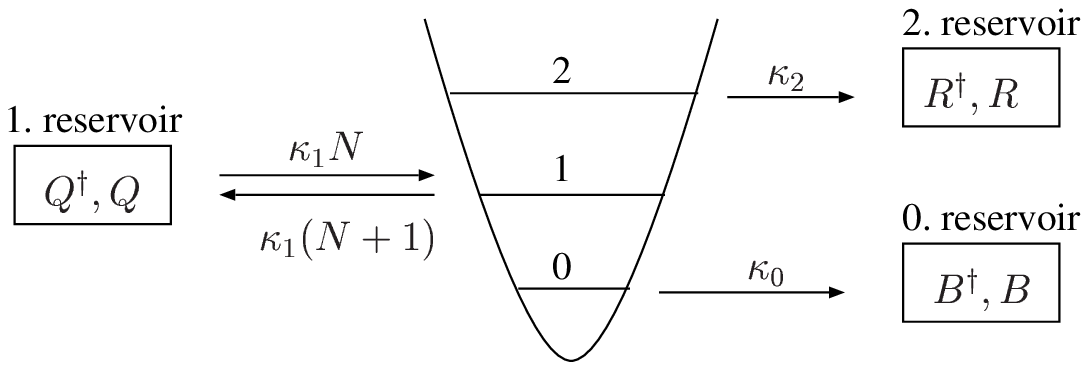}
  \caption{Our simplified model of a cw atom laser consists of a pump
mode 1, the ground state mode 0 which is occupied from the Bose-Einstein
condensate and the weakly bound mode 2 from which the atoms are strongly
coupled out.    }
  \label{fig6}
\end{minipage}\hfill
%
\begin{minipage}[b]{0.46\linewidth}
  \centering\epsfxsize\linewidth\epsffile{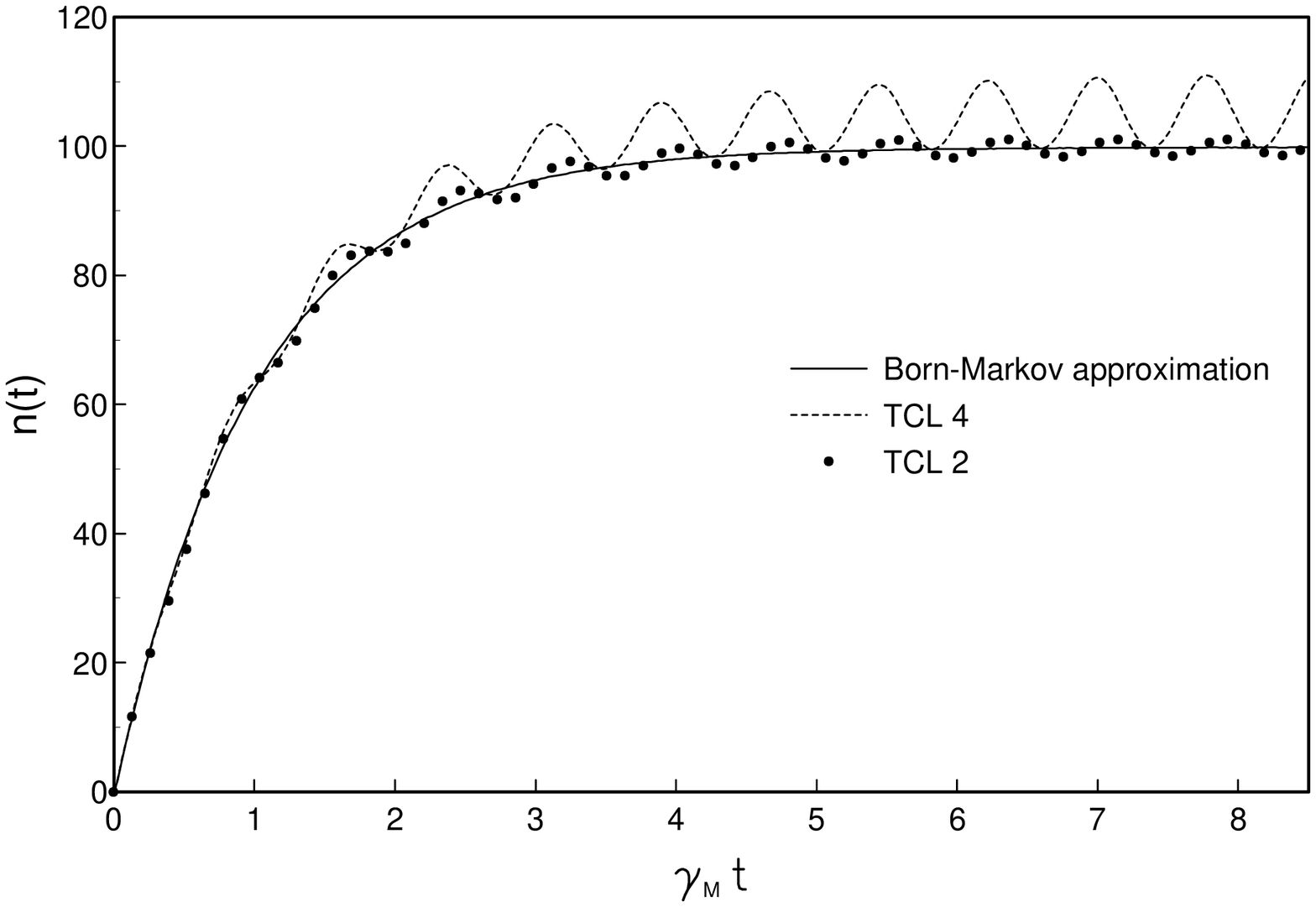}
  \caption{Occupation number $n(t) = \langle a^{\dagger} a  \rangle(t) $ of the BEC in an atomic trap in the case
           of a continuous wave atom laser. Plotted are the
           Born-Markov approximation and the results of a non-Markovian
           perturbative expansion of the output coupling to second and fourth order in the
           coupling strength.}
  \label{fig7}
\end{minipage}
\end{figure}

In Fig.~\ref{fig2} the occupation number of the BEC is plotted for a
coupling strength $\Gamma = 5 \times 10^{4} \;s^{-2}$. While the
Born-Markov approximation fails to render the exact solution, the
depicted 4th order is a very good approximation. Although it is not shown here
the 6th order would be almost perfect.

Fig.~\ref{fig3} shows the results for a coupling strength  $\Gamma =
10^{5} \;\mathrm{s}^{-2}$. Here we clearly see that for realistic
parameters the Born-Markov approximation fails, whereas the
time-convolutionless projection operator technique to 6th order in
the coupling strength reproduces the exact solution. Our numerical
results show that the Born-Markov approximation begins to fail at
a coupling strength $\Gamma \approx 10^{4} \mathrm{s}^{-2}$. This
corresponds to a ratio of the time scales given by
$t_{\text{sys}}/t_{\text{res}} = 80 $. The TCL algorithm to
4th order provides reliable results until about $\Gamma \approx
5 \times 10^{4} \mathrm{s}^{-2}$. For stronger couplings to about $\Gamma
\approx 10^{5} \mathrm{s}^{-2}$ corresponding to a ratio of time scales
$t_{\text{sys}}/t_{\text{res}} = 8$ the 6th order is needed.

For very strong coupling with $\Gamma \approx 10^{6} \mathrm{s}^{-2}$ the
6th order provides reliable results only for short times. The rise in the
atom number after the first collapse shown in
Fig.~\ref{fig4} can not be
reproduced by any algorithm which is based on a perturbative expansion
because the transition rate diverges at this point, and the equation
of motion is no longer analytic \cite{Kappler1}. Nevertheless the
short time behavior is very important for the evaluation of
correlation functions.

Fig.~\ref{fig5} compares the decay rates for the coupling strength
$\Gamma = 10^{5} \mathrm{s}^{-2}$. As expected, the decay rate is
valid for longer times if we include higher order terms in the
perturbation expansion. Although the decay rate in 6th order does
not exactly render the exact solution, it gives a good
approximation over a wide range until about $\gamma_{M} t \approx
4$. It constitutes an essential improvement compared to the
constant Born-Markov decay rate. It is important to emphasize that
the TCL technique is much simpler to perform than the solution of
the exact equation of motion.

\section{Continuous wave atom lasers} \label{cw_al}

The next step after the realization of a pulsed atom laser was to
build a continuous wave (cw) atom laser which was recently
achieved \cite{Hagley,Esslinger}. As in the case of the pulsed
atom laser we model the output coupler through a two-photon Raman
transition. Such an output coupler has recently been
experimentally realized by Hagley {\it et al.} \cite{Hagley}.
Because of the momentum transferred by the Raman transitions this
is the first atom laser with a highly directional output.

In addition to the pulsed atom laser discussed in the first part
we include a pumping mechanism to describe a cw atom laser. In
this section we extend the model of a pulsed atom laser to a pump
mechanism already proposed in \cite{Holland,Zobay,Moore1,Moore2}.
While in these papers the output coupler is treated within the
Born-Markov approximation we will use the same output coupling
mechanism as discussed in section~\ref{pulsed_atom_lasers} to
study the effects of a non-Markovian output coupler. The coupling
to the pump reservoir is a fast process and may therefore be
treated within the Born-Markov approximation.

\subsection{Model of a continuous wave atom laser} \label{pump_m}

We use a binary-collision atom laser scheme based on the near
resonant dipole-dipole interaction. The atomic trap from which we
want to study the output coupling consists of many atomic
modes. Under certain experimental conditions it is possible to
restrict ourselves to three modes of the trap, as discussed in
\cite{Zobay,Moore1}. Atoms of a thermal source - the first
reservoir - are transferred  into the pump mode of the trap. As
shown in Fig.~\ref{fig6} these atoms enter the  trap mode 1 with a
rate $\kappa_{1} N$. They can also leave the trap from this mode
with the rate $\kappa_{1}(1+N)$. The bosonic creation and
annihilation operators of the first reservoir are denoted by
$Q^{\dagger},Q$ and $N$ is the stationary occupation number of the  mode
1 if there were no transitions inside the trap. After the  mode 1
is occupied by more than one atom, two atoms can interact via the
near resonant dipole-dipole interaction through which one atom is
transferred to the weakly bound  mode 2 and the other to the  mode
0 of the atomic trap.

The atoms in the  mode 2 are out-coupled to the second reservoir
with a rate $\kappa_{2}$. To enable evaporative cooling which is
needed to achieve the necessary low temperatures for Bose-Einstein
condensation the condition $\kappa_{2} \gg \kappa_{1}$ must hold.
This means that each atom which is transferred to the highest mode
leaves the trap very fast, such that this mode is most strongly
depleted. Just as in section~\ref{ex_sol} the atoms are
out-coupled through a Raman transition from the  ground state of
the atomic trap. To achieve a steady occupation number of the
ground mode the condition $\kappa_{1} \gg \kappa_{0}$ must hold.
This output coupling to the 0 reservoir will be described taking
into account the non-Markovian character of the dynamics, whereas
the coupling to the other two reservoirs is treated in the
Born-Markov approximation. This is a reasonable assumption because
of the condition $\kappa_{2} \gg \kappa_{1} \gg \kappa_{0}$.

In general we find for the binary-collision atom laser scheme the
Hamiltonian
\begin{equation}
 H = H_{\text{sys}} + V + H_{I,0} + H_{I,1} + H_{I,2}\; ,
\end{equation}
where $H_{\text{sys}}$ is the system Hamiltonian and $V$ describes the
collisions between two atoms. The operators $H_{I,0},H_{I,1},H_{I,2}$ are
responsible for the coupling to the three reservoirs. $H_{\text{sys}}$
and the most general form of the interaction Hamiltonian $V$ can be
written as
\begin{eqnarray}
  H_{\text{sys}} &=& \sum^{2}_{j = 0} \hbar \omega_{j} a^{\dagger}_{j}a_{j}\; , \\
 V &=& \hbar \sum^{2}_{j,k,l,m = 0} g_{j,k,l,m}\; a^{\dagger}_{j} a^{\dagger}_{k} a_{l} a_{m}\;.
\end{eqnarray}
In the interaction Hamiltonian $V$ we do not  consider effects which are negligible at low
densities, as the annihilation of two ground state atoms and the
creation of two atoms in the highest modes. Such transitions are
energetically unfavoured at low densities. Moreover, since $\kappa_{2} \gg \kappa_{1}$
the strongly depleted mode 2 can be eliminated adiabatically~\cite{Holland}. We introduce an
effective reservoir which describes the annihilation of two atoms in the
 mode 1, the creation of an atom in mode 0 and 2, and the
immediate loss of the atom in the mode 2. This effective reservoir
includes the terms describing collision with atoms in the mode 2 which
were included in $V$. With these assumptions the  remaining terms in $V$ are
\begin{eqnarray}
V = \hbar \lbrace  \; g_{0000}\; a^{\dagger}_{0}a^{\dagger}_{0}
a_{0} a_{0} &+& g_{1111} \; a^{\dagger}_{1}a^{\dagger}_{1} a_{1}
a_{1} \nonumber \\ &+&{} g_{0101} \; a^{\dagger}_{0}a_{0}
a^{\dagger}_{1} a_{1}\rbrace\;.
\end{eqnarray}
As shown in \cite{Holland,Zobay} the first two terms in the above
equation cause a broadening of the output spectrum. In this paper
we are only interested in the occupation number of the BEC, and we
will not study  the output spectrum. Because of this we can
ignore these terms since $V$ affects only the off-diagonal terms of
the reduced density matrix $\rho_{\text{sys}}$. Hence we obtain in
the interaction picture
\begin{equation}
H(t) = H_{I,0}(t) + H_{I,1}(t) + H_{\text{eff}}(t),
\end{equation}
where the operators describing the coupling to the three reservoirs
are given by
\begin{mathletters}
\label{int_ham}
\begin{eqnarray} \label{int_ham1}
H_{I,0}(t) &=& - i \hbar \sqrt{\frac{\kappa_{0}}{2}}\; \left( B(t)
a_{0}^{\dagger} e^{i \omega_{0} t} - a_{0} B^{\dagger}(t) e^{-i
\omega_{0} t} \right)\; , \\ H_{I,1}(t) &=&
\sqrt{\frac{\kappa_{1}}{2}}\; \left( Q(t) a_{1}^{\dagger} e^{i
\omega_{1} t} - a_{1} Q^{\dagger}(t) e^{-i \omega_{1} t} \right)\;
,\label{int_ham2} \\ H_{\text{eff}}(t) &=&
\sqrt{\frac{\Omega}{2}}\; \left( R^{\dagger}(t) a_{0}^{\dagger}  a_{1}^{2}
e^{i (\omega_{0}-2 \omega_{1}) t}  \nonumber\right. \\ &
&{}\left.\qquad \qquad -a_{0} a_{1}^{\dagger2} R(t)
e^{-i (\omega_{0}-2 \omega_{1}) t} \right)\;.\label{int_ham3}
\end{eqnarray}
\end{mathletters}
The operators $B,Q,R$ and $B^{\dagger},Q^{\dagger},R^{\dagger}$
are the annihilation and creation operators, respectively, for the
reservoirs shown in Fig.~\ref{fig6}. The  energies of atoms in the
two atomic trap modes 0 and 1 are given by $\hbar \omega_{0}$ and
$\hbar \omega_{1}$. The coupling constant $\Omega$ describes the
strength of the dipole-dipole interaction.

Summarizing, the atomic trap in which the Bose-Einstein condensate is built couples
to three reservoirs. The first reservoir provides the pump mode of
the trap with atoms from a thermal source. Through near resonant
dipole-dipole interaction the atomic trap modes 0 and 2 get
occupied. The immediate loss from atoms in the second
atomic trap mode to the corresponding reservoir is necessary to
enable evaporation cooling. Just as in the case of the pumped atom
laser the Bose-Einstein condensate is coupled out from the mode 0
through a two-photon Raman transition.

\subsection{Derivation of quantum master equations}

In this section we apply the time-convolutionless projection
operator technique to the case of the continuous wave atom laser.
With the help of Eqs. (\ref{entw2}) and (\ref{int_ham}) we obtain to second
order in the coupling strength $\Gamma^{1/2}$
\begin{equation}
 \frac{\partial}{\partial t} \rho_{\text{sys}}(t) = ({\cal L}_{0}^{(2)} + {\cal L}_{\text{out}}^{(2)}
  + {\cal L}_{\text{in}} + {\cal L}_{\text{coll}} )  \rho_{\text{\text{sys}}}(t)\;. \label{bm_mgl_al2}
\end{equation}
The superoperators describing the Lamb shift ${\cal L}_{0}^{(2)}$
and the output coupling ${\cal L}_{\text{out}}^{(2)}$ to second
order are defined through Eqs. (\ref{sup_op_1}). The operators
${\cal L}_{\text{in}}$ and $ {\cal L}_{\text{coll}}$ 
describe the coupling to the thermal reservoir from which the
atoms enter the trap mode 1, and the coupling to the effective
reservoir which enables the evaporative cooling mechanism from the
trap mode 2. These operators are treated within the Born-Markov approximation,
and are thus independent of the expansion parameter. Hence
we have
\begin{mathletters}
\label{sup_op_1}
\begin{eqnarray}
  {\cal L}_{0}^{(n)} R &=&   - \frac{i}{2} S^{(n)}(t)  [a_{0}^{\dagger} a_{0}, R ] \;,\\
  {\cal L}_{\text{out}}^{(n)} R &=& \gamma^{(n)}(t) \left\{ a_{0}\, R\, a_{0}^{\dagger} - \frac{1}{2} a_{0}^{\dagger} a_{0}\, R
   - \frac{1}{2} R\, a_{0}^{\dagger} a_{0} \right\} \;,\\
   {\cal L}_{\text{in}} R &=&   N \kappa_{1} \left\{ a_{1}^{\dagger} R \,a_{1} - \frac{1}{2} a_{1}\, a_{1}^{\dagger}
      R  - \frac{1}{2} R  \,a_{1}\, a_{1}^{\dagger} \right\} \nonumber \\
+  (1&+&N) \kappa_{1} \left\{ a_{1} \,R \, a_{1}^{\dagger} -
\frac{1}{2} a_{1}^{\dagger}\, a_{1}\,
 R - \frac{1}{2} R \,  a_{1}^{\dagger}\, a_{1} \right\} \;, \\
  {\cal L}_{\text{coll}} R &=&  \Omega \; \left\{ a_{0}^{\dagger}  a_{1}^{2}\, R\, a_{1}^{\dagger 2}  a_{0} -
\frac{1}{2} a_{1}^{\dagger 2}  a_{1}^{2}\, a_{0}\, a_{0}^{\dagger} R \right.
\nonumber
\\
    & &{} \left. - \frac{1}{2} R a_{1}^{\dagger 2}  a_{1}^{2}\,  a_{0}\,a_{0}^{\dagger}\right\}\;.
\end{eqnarray}
\end{mathletters}
The functions $\gamma^{(2)}(t)$ and $S^{(2)}(t)$ are given by 
 Eqs. (\ref{s2}) and (\ref{ga_tcl2}). From the above master
equation we obtain the Born-Markov approximation by replacing the
time dependent functions $\gamma^{(2)}(t)$ and $S^{(2)}(t)$
through the Markovian Lamb shift and the Markovian decay rate from
Eq. (\ref{gam_mar}).

In the limit of an atom number in the Bose-Einstein condensate
which is much greater than 1 we get the stationary occupation
number of the ground state within the Born-Markov approximation
\begin{equation}
\langle a^{\dagger}_{0} a_{0} \rangle =  \frac{\kappa_{1}}{2
\gamma_{M}}\;
     \left( N - \frac{1}{2} - \sqrt{\frac{1}{4} +
         \frac{\gamma_{M}}{\Omega}}\; \right)     \;.
\end{equation}
Before solving the Born-Markov master equation and the equation to second order
perturbation theory we evaluate the 4th order term $K^{(4)}$ from
Eq. (\ref{k4}). We obtain after some algebra
\begin{equation} \label{tcl4_eq}
 \frac{\partial}{\partial t} \rho_{\text{sys}}(t) = \left({\cal L}_{0}^{(4)} + {\cal L}_{\text{out}}^{(4)} +
{\cal L}_{\text{in}} + {\cal L}_{\text{coll}} + {\cal
L}_{\text{oc}}\right)  \rho_{\text{\text{sys}}}(t)\;.
\end{equation}
Here, the superoperators ${\cal L}_{0}^{(4)}$ and ${\cal
L}_{\text{out}}^{(4)}$ are defined through Eqs. (\ref{sup_op_1})
and the corresponding functions $\gamma^{(4)}(t)$ and $S^{(4)}(t)$
are given in Eq. (\ref{s4}) and (\ref{g4}), respectively. Naturally the
operators ${\cal L}_{\text{in}}$ and ${\cal L}_{\text{coll}}$ do
not change compared to the second order expansion. In contrast to
the second order perturbative expansion (\ref{bm_mgl_al2}) in the
coupling strength $\Gamma^{1/2}$ the fourth order does not only
change the functions $\gamma^{(n)}(t)$ and $S^{(n)}(t)$ to the
appropriate order, it also adds a new superoperator ${\cal
L}_{\text{oc}}$ to the master equation. ${\cal L}_{\text{oc}}$ is
given by
\begin{eqnarray}
 {\cal L}_{\text{oc}} R &=&
 r(t) \; (\; a_{0}^{\dagger} a_{1}^{2} R a_{0}  a_{1}^{\dagger 2}  -  a_{1}^{2} R a_{0} a_{0}^{\dagger} a_{1}^{\dagger2}\; ) \nonumber \\
& & + r^{*}(t) \;(\; a_{0}^{\dagger} a_{1}^{2} R a_{0}
a_{1}^{\dagger2} - a_{0} a_{0}^{\dagger} a_{1}^{2} R
a_{1}^{\dagger2}\; ) \nonumber
\\ & & + \frac{1}{2}\; r(t) \; (\; a_{0} a_{1}^{\dagger2} a_{1}^{2} R
a_{0}^{\dagger} - a_{0}^{\dagger} a_{0}  a_{1}^{\dagger2}
a_{1}^{2} R \;)\nonumber\\ & & + \frac{1}{2} r^{*}(t) \; (\; a_{0}
R a_{0}^{\dagger} a_{1}^{\dagger2} a_{1}^{2} - R a_{0}^{\dagger}
a_{0} a_{1}^{\dagger2} a_{1}^{2}\; )\;.
\end{eqnarray}
The complex functions $r(t)$ are defined by
\begin{equation}
  r(t) = \Omega \int_{0}^{t} d t_{1} \int_{0}^{t_1} d t_{2}\; f(t_{0}-t_{2})\;.
\end{equation}
The operator ${\cal L}_{\text{oc}}$ appears in the 4th order
perturbation theory due to a mixture of ${\cal L}_{\text{out}}^{(2)}$ and ${\cal L}_{\text{coll}}$.
Note that this master equation is not in Lindblad form.
Eq. (\ref{tcl4_eq}) was solved by integrating the closed system of differential
equations for the diagonal elements of the reduced density operator.

\subsection{Numerical Results}

In our simulations we have chosen the parameters similar to the
pulsed atom laser. The new parameters for the pumping and the
coupling to the effective reservoir are taken to be these of Ref.
\cite{Holland,Zobay}. They are chosen such that the stationary
occupation number $n$ of the ground mode is $n \approx 100$. The
strength of the dipole-dipole coupling is $\Omega = 15\,
\gamma_{M}$ and the coupling constant to the pump reservoir is
$\kappa_{1}=10\, \gamma_{M}$. The parameter $N=20.3$ describes the
stationary occupation number of the pump mode 1 of the atomic
trap, assuming there were no transitions inside the trap.

The only  variable parameter in the simulation is the strength of
the coupling $\Gamma^{1/2}$, which determines the strength of the
coherent output coupling $\gamma_{M}$. This parameter is varied in
our simulations from a value where we expect the Born-Markov
approximation to hold, to parameters where the
time-convolutionless projection operator technique for the pulsed
atom laser was valid in the first part of this paper.

Fig.~\ref{fig7} shows the occupation number of the Bose-Einstein
condensate for a value $\Gamma = 5 \times 10^{4} \mathrm{s}^{-2}$.
For this coupling strength the TCL algorithm agreed very well with
the exact solution in the case of the pulsed atom laser. In the
case of the cw atom laser  we see strong oscillations in the
occupation number of the BEC.  Note also, that in 4th order the mean
occupation number increased compared to the stationary occupation
number resulting from the Born-Markov approximation. 

If we reduce the coupling strength $\Gamma$ to  $\Gamma = 1 \times
10^{4} \mathrm{s}^{-2}$, then there are only very weak
oscillations around the stationary value obtained from the
Born-Markov approximation and the fourth order perturbation theory
agrees well with the second order expansion. In the case of the
pulsed atom laser the Born-Markov approximation also failed for
coupling strengths $\Gamma >  1 \times 10^{4} \mathrm{s}^{-2}$.

As expected this suggests that for the cw atom laser the parameter
regime in which the TCL-algorithm is valid coincides with that of the
pulsed atom laser. Thus the parameters in
Fig.~\ref{fig7} are chosen such that the TCL algorithm to 4th
order should provide reliable results.

The depicted oscillations in the atom number can be interpreted as
an interference effect between two favored transition modes.
The physical origin of this interference phenomena can be seen already
from the second order term $\gamma^{(2)}(t)$ which takes the form
\begin{eqnarray} \label{osci}
 \gamma^{(2)}(t) &=& \int_{0}^{t} ds\; \phi(s) \nonumber \\ &=& 2 \int_{0}^{ \infty}
  d \omega \; J(\omega) \frac{\sin\lbrack (\omega_{0} - \omega)t\rbrack}{\omega_{0} - \omega}\;.
\end{eqnarray}
Here the first factor of the integrand, namely the spectral
density $J(\omega)$, has a $\omega^{-1/2}$ singularity at  $\omega=0$
[see Eq.~(\ref{sp_den})], whereas the second
factor is concentrated around the ground state frequency
$\omega=\omega_{0}$. 
Thus, the main contribution to this integral stems from transitions with
frequencies near $\omega = 0$ and $\omega = \omega_{0}$. The corresponding
transition amplitudes interfere and lead to the observed oscillations
in the decay rate and therefore to the oscillations in the occupation
number.

In a Markovian system, where the spectral density of the coupling
strength $J(\omega)$ is bounded, the second factor of the integral
(\ref{osci}) quickly approaches a $\delta$-function in the limit $t
\rightarrow \infty $, and the oscillations decay on a time scale $t_{\text{res}}$.
This is known as Fermis golden rule.

In contrast, in the case of an atom laser, the contribution near 
$\omega = 0$ is also important for long times. This can be seen by the 
asymptotic behavior of $\gamma^{(2)}(t)$ which can be evaluated
explicitly from Eq. (\ref{osci}) and is given through
\begin{equation}
 \gamma^{(2)}(t) = \gamma_{\text{M}} - \frac{2 \Gamma}{\sqrt{\alpha
     \omega_{0}^{2} t}} \cos(\omega_{0} t + \pi/4)\;.
\end{equation}
According to this relation the decay rate $\gamma^{(2)}(t)$ approaches 
the Markovian rate $\gamma_{M}$ very slowly as $t^{-1/2}$. It is
important to note that this behavior is due to the $\omega^{-1/2}$
singularity of $J(\omega)$ at $\omega=0$ which, in turn, is a direct
consequence of the dispersion relation (\ref{B_formel}) of massive particles.
Thus we observe that non-Markovian effects decay only slowly in time
with an algebraic behavior. This clearly demonstrates that the golden
rule limit $\gamma_{M}$ is relevant only for extremely long times and
shows the importance of a non-Markovian treatment of the atom laser.

From Eq. (\ref{osci}) one clearly sees that the oscillations will
also appear if we abandon the assumption $k_{0}\approx0$. The
function $J(\omega)$ will then be peaked around $\omega_{L} =
\hbar k_{0}^{2}/(2 M)$ but still has a $\omega^{-1/2}$ singularity at
$\omega = 0$. Hence a significant contribution to the integral in
Eq.~(\ref{osci}) comes from $\omega = 0$ and, therefore, the time dependence
of $\gamma^{(2)}(t)$ again shows an oscillatory behavior. These  
oscillations are also present in the case of the damped
Jaynes-Cummings model with detuning (see, e.g., Ref. \cite{Kappler1}), 
but there they are damped exponentially.

\section{Summary}

We have demonstrated that in the case of a pulsed atom laser with
a Raman output coupler in a realistic parameter regime in which
the Born-Markov approximation fails, the time-convolutionless
projection operator technique is able to reproduce the exact
solution.

If we leave the intermediate coupling regime and take the 6th
order terms in the output coupling strength we are able to render
the exact solution for short times to about $t=4/\gamma_{M}$ in
the strong coupling regime. This short time behavior is important
for the evaluation of correlation functions. The long time
behavior after the first collapse of the atom number can not be
reproduced with the TCL algorithm, because the decay rate diverges
at such a point.

In the case of a cw atom laser we have chosen the coupling
strength such that the perturbative expansion for the pulsed atom
laser agreed with the exact solution for the considered times. We found strong oscillations
of the occupation number of the BEC and the mean atom number
increased. 
Our results clearly suggest that these oscillations survive for very long 
times because of the slow algebraic decay of the relaxation rate.

{\bf Acknowledgment}

BK would like to thank the DFG-Graduiertenkolleg { ``Nichtlineare
  Differentialgleichungen''} at the Albert-Ludwigs-Universit\"at
Freiburg for financial support of the research project.

\vspace*{70ex}
\end{multicols}

\widetext

\end{document}